\documentclass[CEJP,PDF]{cej} 
\usepackage{layout}
\usepackage{amsmath}
\usepackage{textcomp}
\usepackage{hyperref}

\usepackage{multirow}
\usepackage{graphicx}
\usepackage{amsmath}

\newcommand{\sqrtsNN}{\mbox{$\sqrt{\mathrm{\it s_{NN}}}$} }
\newcommand{\vtwo}{$v_{2}$ }

\newcommand{\ks}{${K}^{0}_{S}$ }

\newcommand{\lam}{$\Lambda$ }

\def \auau  {Au+Au }

\def \lt {\mbox{$<~$} }

\usepackage{color}

\title{Probe the QCD phase boundary with elliptic flow in relativistic heavy ion collisions at STAR}

\articletype{Editorial}

\author{Shusu Shi\inst{1}$^,$\inst{2} \email{sss@iopp.ccnu.edu.cn} (for the STAR collaboration)
                }

\institute{
     \inst{1} Institute of Particle Physics, Central China Normal University, Wuhan, Hubei, 430079, China
     \inst{2} The Key Laboratory of Quark and Lepton Physics (Central China Normal University), Ministry of Education, Wuhan, Hubei, 430079, China
          }

\abstract{We present measurement of elliptic flow, $v_2$,
for charged and identified particles at midrapidity in Au+Au collisions at \sqrtsNN = 7.7 - 39 GeV.
We compare the inclusive charged hadron $v_2$ to those from
transport model calculations, such as UrQMD model,
AMPT default model and AMPT string-melting model.
We discuss the energy dependence of the difference in $v_2$ between particles and anti-particles.
The $v_2$ of $\phi$ meson is observed to be systematically lower than other particles in
Au+Au collisions at \sqrtsNN = 11.5 GeV.}


\begin{document}
\maketitle


\section{Introduction}
Searching for the region of a possible phase transition between the
Quark Gluon Plasma (QGP) and the hadron gas phase in the QCD phase
diagram is one of the main goals of the Beam Energy Scan (BES) at RHIC.
Due to the sensitivity of underlying dynamics in the early stage of the
collisions, the elliptic flow, $v_2$, could be used as a powerful tool~\cite{review}.
In the top energy (\sqrtsNN = 200 GeV) of RHIC Au+Au and Cu+Cu collisions,
the number of constituent quark (NCQ) scaling in $v_2$ reflects that
the collectivity has been built up at the partonic stage~\cite{starklv2, XiOmega, flowcucu}.
Especially, the NCQ scaling of multi-strange hadrons,
$\phi$ and $\Omega$, provides the clear evidence of partonic collectivity
because they are less sensitive to the late
hadronic interactions~\cite{statphi, phiomega}.
Further, a study based on a multi-phase transport model (AMPT) indicates the NCQ scaling is related to the degrees of freedom in the system~\cite{AMPTNCQ}.
The holding of the NCQ scaling reflects the partonic degree of freedom, whereas the breaking
of the scaling reflects the hadronic degree of freedom.
In reference~\cite{phiBES}, the importance of $\phi$ meson has been
emphasized. Without partonic phase. the $\phi$ meson $v_2$ could be small or zero.
Thus, the measurements of elliptic flow with the Beam Energy Scan data offer us the opportunity to investigate
the phase boundary in the QCD phase diagram.

\begin{figure*}[ht]
\vskip 0cm
\begin{center} \includegraphics[width=0.7\textwidth]{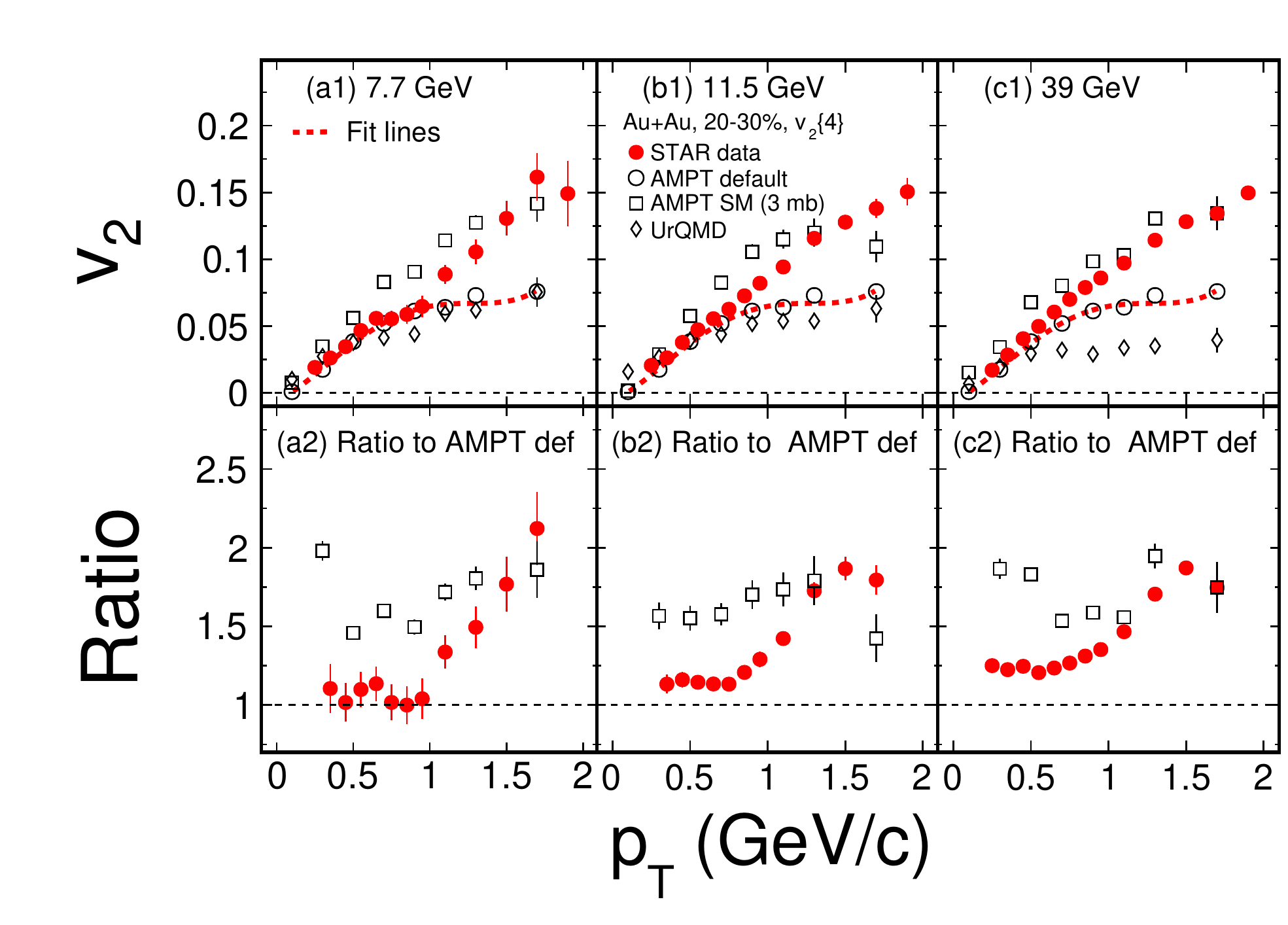}\end{center}
\caption{(Color online) The $v_2\{4\}$ as a function of $p_T$ for $20 - 30\%$ in \auau collisions at \sqrtsNN = 7.7, 11.5 and 39 GeV compared to corresponding results
from UrQMD, AMPT default version and AMPT with string melting version. The bottom panels show the ratios of STAR data and the results of AMPT String-Melting
to the results of AMPT default. Dash lines represent the fit lines of a fifth order polynomial function to the results of AMPT default. }
\label{v2_4_pt_beam_energy}
\end{figure*}

In this paper, we present the \vtwo results of charged and identified hadrons
from the STAR experiment in Au+Au collisions at \sqrtsNN = 7.7 - 39 GeV.
STAR's Time Projection Chamber (TPC)~\cite{STARtpc} is used as the
main detector for event plane determination. The
centrality was determined by the number of tracks from the pseudorapidity region
$|\eta|\le 0.5$. The particle identification for $\pi^{\pm}$, $K^{\pm}$ and $p~(\overline{p})$ is
achieved via the energy loss in the TPC and the time of flight information from the multi-gap resistive plate chamber detector~\cite{STARtof}.
Strange hadrons are reconstructed with the decay channels:
\ks $\rightarrow \pi^{+} + \pi^{-}$, $\phi \rightarrow K^{+} +
K^{-}$, \lam $\rightarrow p + \pi^{-}$ ($\overline{\Lambda}
\rightarrow \overline{p} + \pi^{+}$),
and $\Xi^{-} \rightarrow$ \lam $+\ \pi^{-}$ ($\overline{\Xi}^{+}
\rightarrow$ $\overline{\Lambda}$+\ $\pi^{+}$)). The detailed description of the
procedure can be found in Refs.~\cite{starklv2, XiOmega,klv2_130GeV}.
The event plane method~\cite{v2Methods1} and cumulant method~\cite{cumulant1, cumulant2} are used for the $v_{2}$ measurement.

\section{Results and Discussions}

\begin{figure*}[ht]
\vskip 0cm
\begin{center}\includegraphics[width=0.6\textwidth]{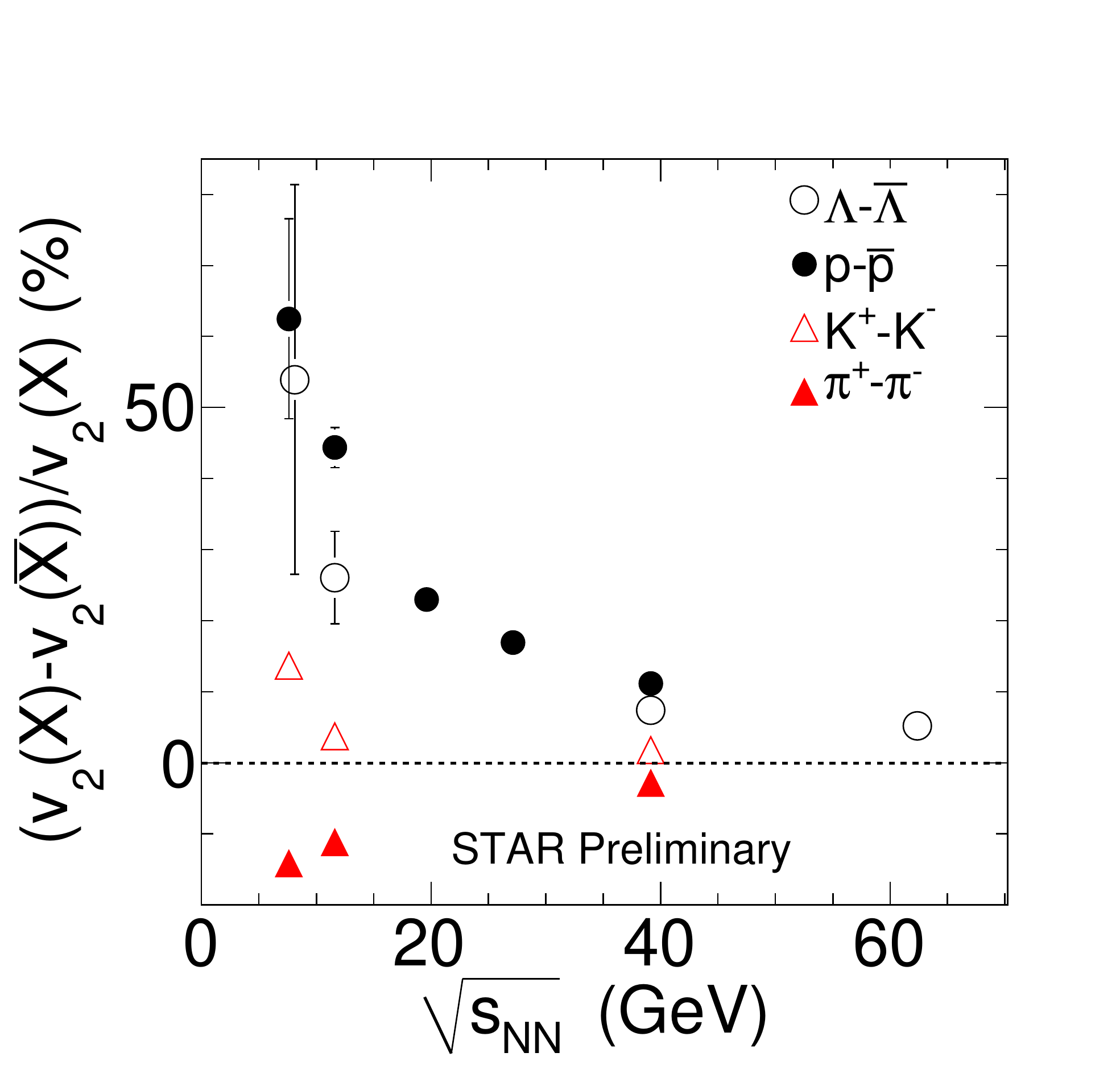}\end{center}
\caption{(Color online) The difference of $v_2$ for particles and anti-particles ($v_{2}(X)-v_{2}(\overline{X})$)
divided by particle $v_2$ ($v_{2}(X)$) as a function of beam energy in Au+Au collisions (0-80\%).}
\label{v2_diff}
\end{figure*}

\begin{figure*}[ht]
\vskip 0cm
\includegraphics[width=1.\textwidth]{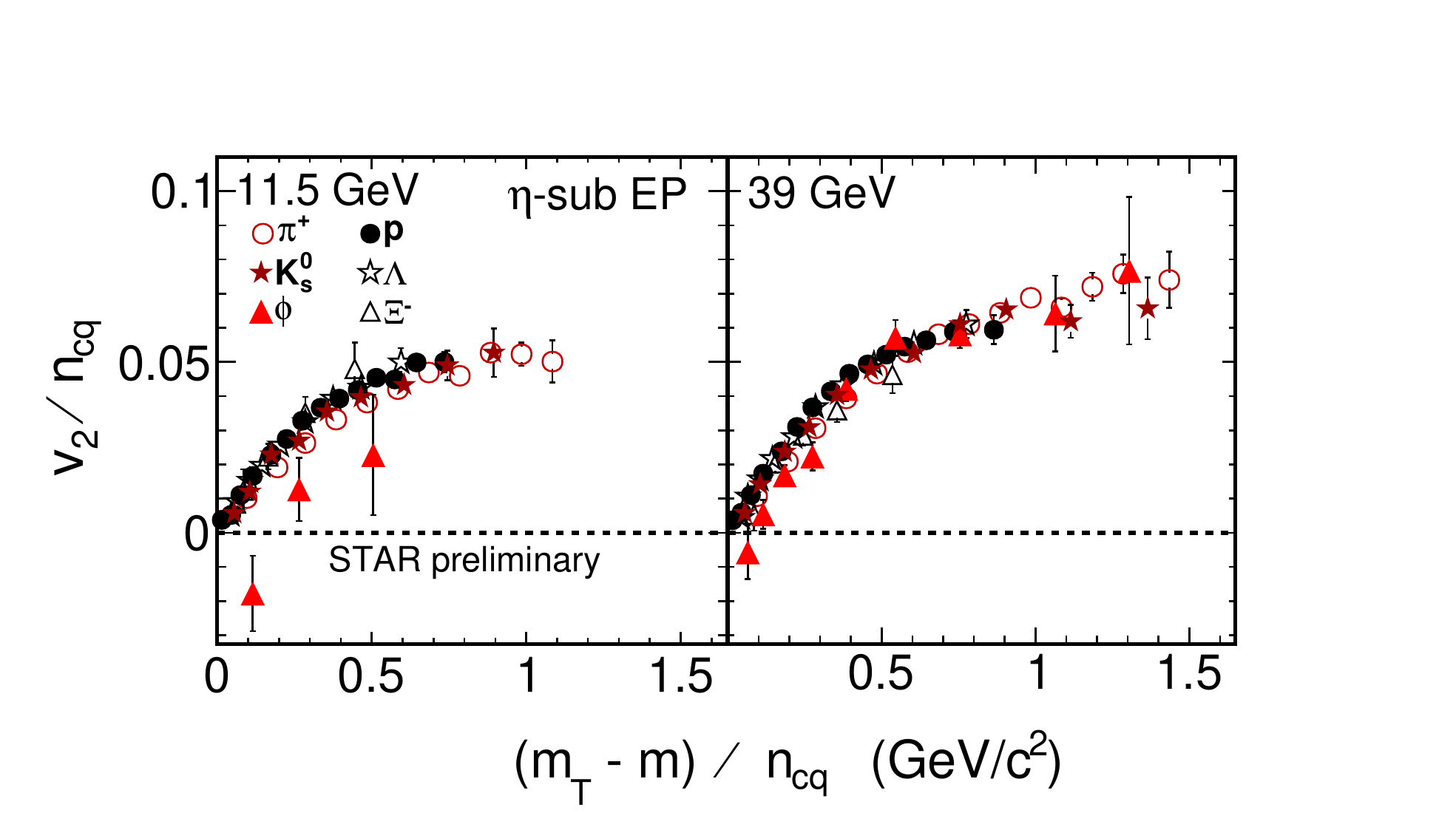}
\caption{(Color online) The number of constituent quark ($n_{\rm cq}$) scaled $v_2$ as a function of transverse kinetic energy over $n_{\rm cq}$
($(m_T - m_0)/n_{\rm cq}$) for various identified particles in Au+Au (0-80\%) collisions at \sqrtsNN = 11.5 and 39 GeV.}
\label{ncq}
\end{figure*}
Figure~\ref{v2_4_pt_beam_energy} shows the results of transverse momentum ($p_T$) dependence of
$v_2\{4\}$ for charged hadrons from  $20-30\%$ centrality class
in Au+Au collisions at \sqrtsNN = 7.7 GeV (a1), 11.5 GeV (b1) and 39 GeV (c1).
To investigate the partonic and hadronic contribution to the final $v_2$ results from different beam energy, transport model calculations from
AMPT~\cite{ampt} and UrQMD~\cite{urqmd} are compared with the STAR data presented. The AMPT default and UrQMD models only take the hadronic
interactions into consideration, while the AMPT String-melting version incorporates both partonic and hadronic interactions. Larger the parton cross section indicates later the hadron cascade starts.
The AMPT String-Melting model with a partonic cross section of 10 mb best describes 200 GeV data; while the AMPT default version falls short by about $40\%$~\cite{ampt}. It suggests that the partonic interactions have to be introduced for the $v_2$ at 200 GeV.
The comparison shows that UrQMD and AMPT model with default setting underpredict the measurements at \sqrtsNN = 39 GeV
for most of the $p_T$ range studied, the differences get reduced as the beam energy decreases.
The data from \auau collisions at \sqrtsNN = 7.7 GeV is pretty close to the results of AMPT default and UrQMD models when $p_T$ is less than 1 GeV/$c$.
For clarity we show the ratios of STAR data and the results of AMPT String-Melting
to the results of AMPT default. The STAR data is closer to the AMPT default and UrQMD models in the lower beam energy.
It indicates the hadronic interactions become more dominant in the lower beam energy.

Figure~\ref{v2_diff} shows the excitation function for the relative difference of $v_2$ between particles and
anti-particles. In order to reduce the non-flow effect, the $\eta$-sub event plane method is used for the measurement.
The $\eta$-sub event plane method is similar to the event plane method, except one
defines the flow vector for each particle based on particles
measured in the opposite hemisphere in pseudorapidity.
An $\eta$ gap of $|\eta| < 0.05$ is used between negative/positive $\eta$ sub-event
to guarantee that non-flow effects are reduced by enlarging the separation
between the correlated particles.
The difference of $v_2$ for baryons is
within 10\% at \sqrtsNN = 39 and 62.4 GeV, while the
difference increases as decreasing beam energy below 39 GeV. At \sqrtsNN = 7.7 GeV,
the difference of protons versus anti-protons is around 60\%. There is no obvious difference for
$\pi^{+}$ versus $\pi^{-}$ (within 3\%) and $K^{+}$ versus $K^{-}$ (within 2\%)
at \sqrtsNN = 39 GeV. As decreasing beam energy,
$\pi^{+}$ versus $\pi^{-}$ and $K^{+}$ versus $K^{-}$ start to show the difference.
The $v_2$ of $\pi^{-}$ is larger than that of
$\pi^{+}$ and the $v_2$ of $K^{+}$ is larger than that of $K^{-}$. This difference between particles and
anti-particles might be due to the baryon transport effect to midrapidity~\cite{transporteffect} or absorption effect in the hadronic stage.
The results could indicate the hadronic interaction become more dominant in lower beam energy.
The immediate consequence of the significant difference between baryon and
anti-baryon $v_2$ is that the NCQ scaling is broken between particles and
anti-particles when \sqrtsNN $\lt$ 39 GeV.
Figure~\ref{ncq} shows the $p_T$ differential $v_2$ for the selected identified particles.
The $v_2$ and $m_{T}-m_{0}$ ($m_{T}=\sqrt{p_{T}^{2}+m_{0}^{2}}$) are divided by number of constituent quark in each hadron.
The similar scaling behavior at \sqrtsNN =200 GeV is observed in Au+Au collisions at \sqrtsNN = 39 GeV. Especially,
the $\phi$ mesons which are not sensitive to the later hadronic interactions follows the same trend of other particles. It suggests that the
partonic degree of freedom and collectivity has been built up at \sqrtsNN = 39 GeV.
Whereas, at \sqrtsNN = 11.5 GeV, the $v_2$ for $\phi$ mesons falls off from other
particles. The mean deviation to the $v_2$ of pions is 2.6 $\sigma$.
It indicates that the hadronic interactions are dominant at \sqrtsNN = 11.5 GeV.

\section{Summary}
In summary, we present the $v_2$ measurements for charged hadrons and identified hadrons in Au+Au collisions at
\sqrtsNN= 7.7 - 39 GeV. Comparison of charged hadron $v_2$ is made with transport model calculations
show agreement between data, UrQMD and AMPT models decreases as beam energy increases.
The data at lower beam energy is closer to AMPT default and UrQMD models. The comparison suggests the hadronic interactions are more dominant in the lower beam energy.
The difference between the $v_2$ of particles and anti-particles is observed.
The baryon and anti-baryon $v_2$ show significant difference in \sqrtsNN $\lt$ 39 GeV.
The difference of $v_2$ between
difference particles and anti-particles (pions, kaons, protons and $\Lambda$s)
increases with decreasing of the beam energy. The $v_2$ of $\phi$ meson falls off from other particles
at \sqrtsNN = 11.5 GeV. Experimental data indicates the hadronic interactions are dominant when
\sqrtsNN $\leq$ 11.5 GeV.

\section{Acknowledgments}
This work was supported in part by the National Natural Science Foundation of China under grant No. 11105060, 10775060 and 11135011.


\begin{thebibliography}{00}
\bibitem{review} S. A. Voloshin, A. M. Poskanzer and R. Snellings, arXiv:0809.2949.
\bibitem{starklv2} J. Adams {\it et al.} (STAR Collaboration), Phys. Rev. Lett. {\bf 92}, 052302 (2004).
\bibitem{XiOmega} J. Adams {\it et al.} (STAR Collaboration), Phys. Rev. Lett. {\bf 95}, 122301 (2005).

\bibitem{flowcucu} B. I. Abelev {\it et al.}, (STAR Collaboration), Phys. Rev. {\bf C 81}, 044902 (2010).
\bibitem{statphi} B. I. Abelev {\it et al.}, (STAR Collaboration), Phys. Rev. Lett. {\bf 99} 112301 (2007).
\bibitem{phiomega} S. S. Shi (for the STAR collaboration), Nucl. Phys. A {\bf 830}, 187c (2009).
\bibitem{AMPTNCQ} F. Liu, K.J. Wu, and N. Xu, J. Phys. G {\bf 37} 094029(2010).
\bibitem{phiBES}B. Mohanty and N. Xu, J. Phys. G {\bf 36}, 064022(2009).
\bibitem{STARtpc} K. H. Ackermann {\it et al.} (STAR Collaboration), Nucl. Instrum. Methods A {\bf 499}, 624 (2003).
\bibitem{STARtof} W. J. Llope (STAR TOF Group), Nucl. Instr. and Meth. B {\bf 241}, 306 (2005).
\bibitem{klv2_130GeV} C. Adler {\it et al.} (STAR Collaboration), Phys. Rev. Lett. {\bf 89}, 132301 (2002).
\bibitem{v2Methods1} A. M. Poskanzer and S. A. Voloshin, Phys. Rev. C \textbf{58} 1671 (1998).
\bibitem{cumulant1}N. Borghini, P. M. Dinh, and J.-Y. Ollitrault, Phys. Rev. C \textbf{63},
054906 (2001).
\bibitem{cumulant2}N. Borghini, P. M. Dinh, and J.-Y. Ollitrault, Phys. Rev. C \textbf{64},
054901 (2001).
\bibitem{ampt} Z. Lin {\it et al.} , Phys. Rev. C {\bf 68}, 054904 (2003).
\bibitem{urqmd} H. Petersen {\it et al.}, arXiv: 0805.0567v1.
\bibitem{transporteffect} J. Dunlop, M.A. Lisa and P. Sorensen, arXiv:1107.3078.

\end{thebibliography}
\end{document}